# The EU Digital Services Act: what does it mean for online advertising and adtech?

## Pieter Wolters & Frederik Zuiderveen Borgesius

*Draft. We would love to hear your comments or suggestions. Please contact us before citing. Thank you!*

*pieter.wolters@ru.nl & frederikzb@cs.ru.nl*

*iHub, Radboud University, The Netherlands*

***Abstract*** What does the Digital Services Act (DSA) mean for online advertising? We describe and analyse the DSA rules that are most relevant for online advertising and adtech (advertising technology). We also highlight to what extent the DSA's advertising rules add something to the rules in the General Data Protection Regulation (GDPR) and the ePrivacy Directive. The DSA introduces several specific requirements for online advertising. First, the DSA imposes transparency requirements in relation to advertisements. Second, very large online platforms (VLOPs) should develop a publicly available repository with information about the ads they presented. Third, the DSA bans profiling-based advertising (behavioural advertising) if it uses sensitive data or if it targets children.

Besides these specific provisions, the general rules of the DSA on illegal content also apply to advertising. Advertisements are a form of information, and thus subject to the general DSA rules. Moreover, we conclude that the DSA applies to some types of ad tech companies. For example, ad networks, companies that connect advertisers to publishers of apps and websites, should be considered platforms. Some ad networks may even qualify as VLOPs.

Hence, ad networks must comply with the more general obligations in the DSA. The application of these general rules to advertisements and ad networks can have far-reaching effects that have been underexplored and deserve further research. We also show that certain aspects of the DSA are still unclear. For instance, we encourage the European Commission or regulators to clarify the concepts of 'online platform' and 'recipients' in the context of ad networks and other adtech companies.

## Table of Contents

# 1 Introduction

Online advertising is among the main income streams for some of the largest companies in the world, such as Facebook (Meta) and Google (Alphabet).[1] Such companies often personalise ads based on the interests of the users. To this end, vast amounts of information about hundreds of millions of people are collected. Many people see this as a privacy interference.[2] Moreover, advertising may contain fraud, misinformation and other illegal or harmful content. Online advertising is often offered through online platforms that connect their users to advertisements supplied by third parties.

This paper looks at the obligations of online platforms in relation to online advertising that follow from the Digital Services Act (DSA). The DSA is an EU regulation that applies from February 2024.[3] It includes general rules for intermediary services, a

The authors thank Sarah Eskens, Lex Zard, and the participants to the PLSC session at Digital Legal Talks 2025 for their valuable comments.

[1] See e.g. Alphabet, 'Alphabet Announces First Quarter 2024 Results' (25 April 2024), <https://abc.xyz/assets/91/b3/3f9213d14ce3ae27e1038e01a0e0/2024q1-alphabet-earnings-release-pdf.pdf> accessed 20 December 2024; Meta, 'Meta Reports First Quarter 2024 Results' (24 April 2024), <https://s21.q4cdn.com/399680738/files/doc_financials/2024/q1/Meta-03-31-2024-Exhibit-99-1_FINAL.pdf> accessed 20 December 2024. See generally on the economics of online advertising: Choi, H., Mela, C. F., Balseiro, S. R., & Leary, A. (2020). Online display advertising markets: A literature review and future directions. *Information Systems Research*, *31*(2), 556-575; MacKenzie, D., Caliskan, K., & Rommerskirchen, C. (2023). The longest second: Header bidding and the material politics of online advertising. *Economy and Society*, *52*(3), 554-578.
[2] S. C. Boerman, S. Kruikemeier, and F. J. Zuiderveen Borgesius, 'Online behavioral advertising: a literature review and research agenda', Journal of Advertising, 2017-46-3, p. 363-376, 2017.
[3] Regulation (EU) 2022/2065 of the European Parliament and of the Council of 19 October 2022 on a Single Market for Digital Services and amending Directive 2000/31/EC (Digital Services Act) [2022] OJ L277/1, art 93(2). See also art 92 and Section 2.2 about the earlier application to providers of very large online platforms and very large online search engines.

concept that includes various actors involved in online advertising, but also more specific rules in relation to online advertising. The paper explores the following question: what does the DSA mean for online advertising?

We describe and analyse the DSA rules that are most relevant for online advertising and highlight to what extent they add something to the rules in the General Data Protection Regulation (GDPR)[4] and the ePrivacy Directive.[5] Some topics are outside the scope of this paper. We do not compare the DSA with more specific content moderation rules from other regulation that apply to online advertising. For example, we do not discuss the 'Political advertising regulation'[6] or the 'Product safety pledge'[7] in relation to advertisements for dangerous goods. The DSA's provision on dark patterns is also outside the scope of this paper.[8]

The paper is structured as follows. Section 2 briefly introduces the DSA. Next, we turn to the DSA's rules on online advertising. Section 3 discusses the relevant definitions in the DSA, which set the scope of the rules on advertising and commercial communications. Section 4 highlights the relevance of the *general* rules of the DSA for online advertising. Next, we focus on the more specific rules in the DSA. Section 5 discusses transparency requirements for recommender systems and individual advertisements. Section 6 discusses reporting obligations and advertising repositories. Section 7 discusses the DSA's restrictions on profiling. Section 8 concludes.

---

[4] Regulation 2016/679/EU of the European Parliament and of the Council of 27 April 2016 on the protection of natural persons with regard to the processing of personal data and on the free movement of such data [2016] OJ L119/1.

[5] Directive 2002/58/EC of the European Parliament and of the Council of 12 July 2002 concerning the processing of personal data and the protection of privacy in the electronic communications sector (Directive on privacy and electronic communications), 2002 O.J. (L 201), last updated by Directive 2009/136/EC of the European parliament and of the Council of 25 November 2009.

[6] Regulation (EU) 2024/900 of the European Parliament and of the Council of 13 March 2024 on the transparency and targeting of political advertising [2024] OJ L900.

[7] <https://commission.europa.eu/business-economy-euro/doing-business-eu/eu-product-safety-and-labelling/product-safety/product-safety-pledge_en> accessed 20 December 2024.

[8] DSA, art 25. See on that provision: Cristiana Santos, Nataliia Bielova, Sanju Ahuja, Christine Utz, Colin Gray, and Gilles Mertens, 'Which Online Platforms and Dark Patterns Should Be Regulated under Article 25 of the DSA?, pre-print, 24 July 2024, <https://papers.ssrn.com/sol3/papers.cfm?abstract_id=4899559> accessed 20 December 2024.

The DSA and its implications for online advertising have been the focus of various academic studies. However, these studies have not covered the full breadth of its relevance. The previous studies have mostly focused on specific provisions. [9] Furthermore, their focus is on the platforms on which the advertisements are displayed, such as the social media or video platforms, and the influencers that are active on them.[10]

In contrast, we argue that the influence of the DSA is much broader. We argue that the general rules regarding content moderation also present important ramifications for online advertising. Furthermore, we show that the relevance of the DSA for online advertising is not limited to social media and other 'typical' types of platforms. Instead, the DSA also applies to other forms of advertising or 'adtech' companies such as ad networks.

## 2    General introduction to the DSA

This Section provides a general introduction of the aim and context (Section 2.1), the layered approach (Section 2.2) and enforcement (Section 2.3) of the DSA. The application of the various concepts and obligations in the context of online advertising is further discussed in the subsequent Sections.

### 2.1  The aim and context of the DSA

The aim of the DSA is to contribute to the proper functioning of the internal market for intermediary services.[11] To this end, the DSA applies to intermediary services that are offered to recipients in the EU, regardless of where the provider is based.[12] Hence, the DSA can also apply to, for instance, companies from the US and China. The DSA creates 'content moderation' rules that are designed to strike a balance between various competing interests.[13] On the one hand, the DSA is designed to foster a safe, predictable and trusted online environment. To this end, the DSA contains obligations to prevent the dissemination of illegal and harmful content.[14] On the other hand, the DSA should also protect fundamental rights such as freedom of expression and information. The various obligations are therefore formulated cautiously and combined with safeguards to prevent the providers of intermediary services from arbitrarily removing permissible content.[15] Moreover, a conditional exemption from liability means that providers have less incentives to remove content out of fear of liability.[16]

---

[9] E.g. Sebastian Becker and Jan Penfrat, 'The DSA Fails to Reign in the Most Harmful Digital Platform Businesses – But It Is Still Useful' in Joris van Hoboken and others, *Putting the DSA into Practice* (Verfassungsblog 2023) 57-59; Bram Duivenvoorde and Catalina Goanta, 'The regulation of digital advertising under the DSA: A critical assessment' (2023) 51 Computer Law & Security Review 105870, 8-11.
[10] Duivenvoorde and Goanta (n 7) 6-8.
[11] See generally on the DSA: Martin Husovec, *Principles of the Digital Services Act* (Oxford University Press 2024).

This general tension between competing interests also determines the DSA's treatment of online advertising. The EU aims to prevent dissemination of illegal and harmful advertisements. At the same time, safeguards should limit the arbitrary removal of permissible advertisements to protect fundamental rights such as freedom of expression and information.

The DSA is a general instrument. It applies to all intermediary services and all forms of content, including advertisements (see also Section 3). Other legal instruments, including codes of conduct and other forms of soft law, add more specific content moderation rules for specific types of content and specific services.[17] These rules can also be relevant for advertisements. For example, the 'Copyright in the Digital Single Market Directive' imposes rules for 'online content-sharing service providers' in relation to copyright-protected works.[18] Next, the 'Memorandum of understanding' on the sale of counterfeit goods via the internet' is meant to prevent the sale of counterfeit goods.[19] Similarly, the 'Product Safety Pledge' is designed to restrict products that do not comply with product safety requirements.[20] Finally, the 'Strengthened Code of Practice on Disinformation' and the 'Political advertising regulation' contain additional and more detailed rules on political advertisements.[21]

---

[12] DSA, art 2(1).

[13] About the aims of the DSA in general, see also DSA, recitals 3, 4, 40, art 1(1). For an overview of the various interests, see Pieter Wolters and Raphaël Gellert, 'Towards a better notice and action mechanism in the DSA' (2023) 14 JIPITEC 403, 405-408. The phrase 'Content moderation' is defined in DSA, art 3(t).

[14] DSA, recitals 9, 70, 80, art 1(2)(a), 34(1)(a), (2).

[15] DSA, recitals 3, 4, 50, 52, art 1(1).

[16] DSA, art 1(2)(a); Joined Cases C-682/18 and C-683/18 *Youtube* [2020] ECLI:EU:C:2020:586, Opinion of AG Saugmandsgaard Øe, para 189; European Commission, 'Impact assessment Accompanying the Proposal for a Regulation of the European Parliament and of the Council on a Single Market For Digital Services (Digital Services Act) and amending Directive 2000/31/EC PART 1/2' SWD (2020) 348 final, box 1; Raphaël Gellert and Pieter Wolters, *The revision of the European framework for the liability and responsibilities of hosting service providers* (Report for the Dutch Ministry of Economic Affairs and Climate Policy, 2021) 21-23 (with references to further literature).

[17] For a more extensive overview of the obligations of hosting service providers, see e.g. Gellert and Wolters (n 14) 34-51.

Other European legal instruments contribute to the proper functioning of the internal market for intermediary services through other means. Most notably, the 'Digital Markets Act' ('DMA') is designed to ensure fair competition in markets where large 'gatekeepers' providing 'core platform services' are present.[22] Other instruments such as the 'eCommerce Directive' and the 'P2B Regulation' aim to improve the online internal market through various transparency obligations.[23] In this paper, however, we focus on the DSA.

---

[18] Directive (EU) 2019/790 of the European Parliament and of the Council of 17 April 2019 on copyright and related rights in the Digital Single Market and amending Directives 96/9/EC and 2001/29/EC [2019] OJ L130/92 (CDSMD), art 1, 2(6).
[19] <https://single-market-economy.ec.europa.eu/industry/strategy/intellectual-property/enforcement-intellectual-property-rights/memorandum-understanding-sale-counterfeit-goods-internet_en> accessed 3 June 2024.
[20] Product Safety Pledge: Voluntary Commitment of Online Marketplaces with Respect to the Safety of Non- Food Consumer Products Sold Online by Third Party Sellers (2018) <https://commission.europa.eu/business-economy-euro/product-safety-and-requirements/product-safety/product-safety-pledge_en> accessed 3 June 2024.
[21] The Strengthened Code of Practice on Disinformation 2022 <https://digital-strategy.ec.europa.eu/en/library/2022-strengthened-code-practice-disinformation> accessed 20 December 2024, commitments 5-9; Regulation (EU) 2024/900 of the European Parliament and of the Council of 13 March 2024 on the transparency and targeting of political advertising [2024] OJ L2024/900. About these rules and their relation to advertisements, see e.g. Paolo Cavaliere, 'The Truth in Fake News: How Disinformation Laws Are Reframing the Concepts of Truth and Accuracy on Digital Platforms' (2022) 3 European Convention on Human Rights Law Review 481, 489-505; Max Zeno van Drunen, Natalie Helberger and Ronan Ó Fathaigh, 'The beginning of EU political advertising law: unifying democratic visions through the internal market', (2022) 30 International Journal of Law and Information Technology 181, 192-199.
[22] Regulation (EU) 2022/1925 of the European Parliament and of the Council of 14 September 2022 on contestable and fair markets in the digital sector and amending Directives (EU) 2019/1937 and (EU) 2020/1828 (Digital Markets Act) [2022] OJ L265/1, art 1(1), 2(1), (2), 3.
[23] Directive 2000/31/EC of the European Parliament and of the Council of 8 June 2000 on certain legal aspects of information society services, in particular electronic commerce, in the Internal Market (Directive on electronic commerce) [2000] OJ L178/1, art 1(1), 5-7, 10; Regulation (EU) 2019/1150 of

## 2.2 The layered approach of the DSA

Chapter III of the DSA imposes obligations on the providers of 'intermediary services'. The chapter has a layered approach: some rules apply to all providers of intermediary services, whereas others only apply to certain categories or even subcategories of intermediary services.

Section 1 imposes obligations that apply to all providers of intermediary services. Pursuant to Article 3(g) DSA, this concept refers to 'mere conduit', 'caching' and 'hosting' services. According to Article 3(j) and recital 28 DSA, an 'online search engine' is also an intermediary service.[24]

Section 2 contains additional obligations for the providers of hosting services. Pursuant to Article 3(g)(iii) DSA, a hosting service is defined as a service 'consisting of the storage of information provided by, and at the request of, a recipient of the service'. The 'recipient' is, roughly speaking, the user. The DSA defines the 'recipient of the service' as 'any natural or legal person who uses an intermediary service, in particular for the purposes of seeking information or making it accessible.'[25] The definition thus

---

the European Parliament and of the Council of 20 June 2019 on promoting fairness and transparency for business users of online intermediation services [2019] OJ L186/57, art 1(1).

[24] However, the DSA does not clarify whether 'online search engine' should be considered as one of the services mentioned in Article 3(g) or whether it constitutes a fourth category. This is especially relevant in order to determine whether a search engine can benefit from one of the exceptions from liability of Articles 4 through 6 DSA. Furthermore, the qualification as a hosting service would lead to the additional obligations of Section 2. Cf Council of the European Union, 'Proposal for a Regulation of the European Parliament and of the Council on a Single Market For Digital Services (Digital Services Act) and amending Directive 2000/31/EC - General approach' 13203/21, art 2(f), 4(1). The Council proposed to make explicit that search engines can benefit from the same exemption as caching services. Much has been written about the qualification of search engines and whether they can benefit from the exemptions. See e.g. Lilian Edwards, *Role and Responsibility of Internet Intermediaries in the Field of Copyright and Related Rights* (WIPO 2011) 13-14; Joris V.J. van Hoboken, *Search engine freedom: on the implications of the right to freedom of expression for the legal governance of Web search engines* (UvA-DARE 2012) 227-230; Christina J. Angelopoulos, *European intermediary liability in copyright: A tort-based analysis* (UvA-DARE 2016) 46-47; Giovanni Sartor, *Providers Liability: From the Commerce Directive to the future* (in-depth analysis for the European Parliament, PE 614.179, 2017) 21-29; Pieter T.J. Wolters, 'Search Engines, Digitalization and National Private Law' [2020] ERPL 795, 811.

[25] DSA, art 3(b).

also includes business users that use the service to disseminate information.[26] Hence, an advertiser can also be a recipient.[27]

Section 3 provides further obligations for 'online platforms', with the exception of micro and small enterprises.[28] 'Online platform' is a subcategory of hosting services. Online platforms do not only store the provided information. They also 'disseminate' it to a potentially unlimited number of third parties.[29] This definition includes social media and online marketplaces that make information easily accessible without requiring further action by the recipient that provided the content.[30] The definition of online platforms excludes services such as cloud computing or web hosting. Although information stored on these hosting services can also be disseminated, this requires further action by the recipient, as recipients have to actively share the access or address of the content.

Sections 4 and 5 of Chapter III impose obligations on two subcategories of online platforms. Section 4 applies to online marketplaces that facilitate to business-to-consumer transactions or, in the terminology of the DSA, 'online platforms allowing consumers to conclude distance contracts with traders'.[31] Micro and small enterprises are again exempted.[32]

---

[26] See also DSA, recital 2.
[27] See also explicitly 'Questions and Answers on identification and counting of active recipients of the service under the Digital Services Act', point 8 <https://digital-strategy.ec.europa.eu/en/library/dsa-guidance-requirement-publish-user-numbers> accessed 20 December 2024.
[28] DSA, art 19(1). Micro and small enterprises are also excluded from the reporting obligation of Article 15. DSA, art 15(2). Micro and small enterprises employ fewer than 50 persons and have an annual turnover and/or annual balance sheet total that does not exceed EUR 10 million. Commission Recommendation 2003/361/EC of 6 May 2003 concerning the definition of micro, small and medium-sized enterprises [2003] OJ L124/36, Annex, art 2(2). The exclusion does not apply to 'very large online platforms'. DSA, art 19(2).
[29] DSA, art 3(i), (k).
[30] DSA, recitals 13, 14.
[31] Cf the definition of 'online marketplace' the Modernisation Directive, which applies to both business-to-consumer and consumer-to-consumer transactions. Directive (EU) 2019/2161 of the European Parliament and of the Council of 27 November 2019 amending Council Directive 93/13/EEC and Directives 98/6/EC, 2005/29/EC and 2011/83/EU of the European Parliament and of the Council as regards the better enforcement and modernisation of Union consumer protection rules [2019] OJ L 328/7, art 3(1)(b), 4(1)(e).
[32] DSA, art 29.

Finally, Section 5 imposes additional obligations on 'very large online platforms' ('VLOPs') and 'very large online search engines' ('VLOSEs') with more than 45 million monthly 'active recipients', as defined in Article 3(p) DSA, in the European Union.[33] For them, the DSA already applied four months after their designation as such by the European Commission.[34] As the first VLOPs and VLOSEs have been designated as such in April 2023, they were bound by the DSA from August 2023, almost half a year before the DSA's general application from 17 February 2024.[35] Examples of very large platforms are Amazon, Google, LinkedIn, Meta, Twitter, and Zalando.[36]

Figure 1 gives an overview of the layered approach of the DSA.

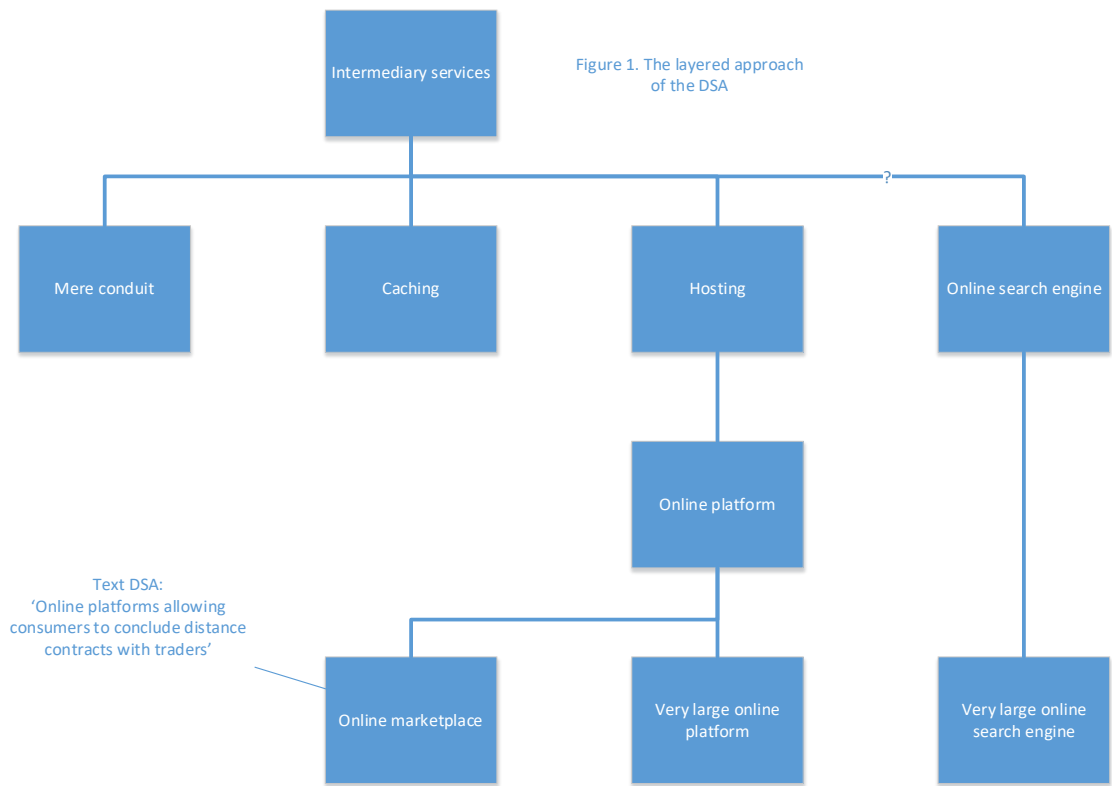

Figure 1. The layered approach of the DSA

---

### 2.3 The enforcement of the DSA

Chapter IV provides the rules on the enforcement of the DSA. Chapter IV places the emphasis on supervision by (cooperating)[37] national 'competent authorities', including a 'Digital Service Coordinator' with various powers of investigation and enforcement.[38] The member states should lay down 'effective, proportionate and dissuasive' penalties,[39] including fines with a maximum of 6% of the annual worldwide turnover of the provider of the intermediary service. Periodic penalty payments can be up to 5% of the average daily worldwide turnover.[40]

The enforcement by the competent authorities is complicated by the international character of many intermediary services. To avoid conflicting investigations by several national authorities, the member state in which the main establishment of the provider of the intermediary service is located has exclusive powers to supervise and enforce the DSA.[41] However, the European Commission is responsible for the supervision of VLOPs and VLOSEs. This rule aims to prevent disproportionate burdens and dependence on the member states in which many of the largest providers are located (in other words: Ireland).[42]

In addition to the extensive rules on enforcement by competent authorities, the DSA also contains several provisions on enforcement by private actors. The recipients of the intermediary services have the right to lodge a complaint with the competent authorities.[43] They may also claim compensation if they suffer damage due to a violation of the DSA.[44] Finally, the DSA creates several redress possibilities in relation to content moderation decisions (Section 4). These possibilities do not prevent a recipient from initiating proceedings before a court or alternative dispute resolution procedure.[45]

---

[37] See DSA, art 57-63.
[38] DSA, art 49-51.
[39] DSA, art 52(2).
[40] DSA, art 52, 74, 76.
[41] DSA, art 56
[42] See also DSA, recital 124.
[43] DSA, art 53.
[44] DSA, art 54.
[45] DSA, art 21(1), (9).

# 3    The applicability of the DSA to online advertisements

## 3.1  Advertisements and related forms of online content

The DSA contains several rules that apply to advertising. Advertisements are a form of 'information' or 'content' that are subject to the general content moderation rules of the DSA.

'Information' and 'content' are not defined in the DSA. However, their meaning can be inferred from the description of the concept of 'illegal content'. Illegal content is, in short, 'any information that, in itself or in relation to an activity, including the sale of products or the provision of services, is not in compliance with' EU or member state law.[46] Since *any type* of information can be illegal, the form is not important. Illegal content can be a video or picture, but also a message that constitutes stalking.[47] Furthermore, content can also refer to activities that are illegal. For example, an offer to sell perfectly legal products can be illegal content if this offer violates consumer law.[48] An 'advertisement' is thus a type of 'information' that can also be 'illegal content'. The next section of this paper discusses several DSA rules on illegal content.

The DSA defines 'advertisement' as follows: 'information designed to promote the message of a legal or natural person, irrespective of whether to achieve commercial or non-commercial purposes, and presented by an online platform on its online interface against remuneration specifically for promoting that information'.[49] The definition also applies to advertising messages without a commercial purpose.[50] Therefore, 'political' advertising by, for instance, governments, political parties, and non-governmental organisations also falls within the DSA's definition of 'advertisement'. In this light, it takes a different approach than the Misleading and Comparative Advertising Directive.[51] That Directive is limited to advertisements with a commercial purpose.[52]

---

[46] DSA, art 3(h).
[47] DSA, recital 12.
[48] DSA, recital 12.
[49] DSA, art 3(r).
[50] See also Duivenvoorde and Goanta (n 7) 5.

Another aspect of the advertising definition is that the information is 'shown against remuneration specifically for promoting that information'.[53] Therefore, some marketing messages fall outside the scope of 'advertising'. For example, a business may also have a 'normal' account on an online platform that it uses to promote its products. The content that is disseminated by this account may be shown to or shared by other recipients. Such 'organic' content falls outside the scope of 'advertising' if the company does not pay to promote this information.

Another category of online marketing also falls outside the scope of the DSA's advertising definition: influencer marketing.[54] In the words of Goanta and Ranchordás, 'modern influencer marketing is a digital form of word-of-mouth ('WOM') advertising'.[55] For example, a brand could pay a popular video blogger (an influencer) to mention the brand in YouTube videos. In such a case, the brand does not pay the platform for the advertising, and the marketing falls outside the DSA's advertising definition.

The previous examples show that not all forms of online marketing are covered by the DSA's definition of 'advertisement'. However, the DSA also includes specific rules for 'commercial communication'. The DSA defines this concept by referring to the definition in the eCommerce Directive.[56] That directive defines 'commercial communication' as 'any form of communication designed to promote, directly or indirectly, the goods, services or image of a company, organisation or person pursuing a commercial, industrial or craft activity or exercising a regulated profession. (…)'.[57]

---

[51] Directive 2006/114/EC of the European Parliament and of the Council of 12 December 2006 concerning misleading and comparative advertising (codified version) [2006] OJ L 376/21. ,

[52] That Directive defines advertising as 'the making of a representation in any form *in connection with a trade, business, craft or profession in order to promote the supply of goods or services*, including immovable property, rights and obligations'. Article 2(a) of the Misleading and Comparative Advertising Directive 2006/114/EC, emphasis added. A much-cited paper in the field of marketing defines 'advertising' as follows: 'paid, mediated form of communication from an identifiable source, designed to persuade the receiver to take some action, now or in the future'. Curran CM and Richards JI, 'Oracles on 'advertising': Searching for a definition' (2002) 31(2) Journal of Advertising 63.

[53] DSA, art 3(r).

[54] Duivenvoorde and Goanta (n 7) 6.

[55] Catalina Goanta and Sofia Ranchordás, 'The Regulation of Social Media Influencers: An Introduction', in *The Regulation of Social Media Influencers* (Edward Elgar Publishing 2020), 1-22. https://doi.org/10.4337/9781788978286.00008

[56] DSA, art 3(w): ''commercial communication' means 'commercial communication' as defined in Article 2, point (f), of Directive 2000/31/EC.

Hence, unlike the concept 'advertisement' in the DSA, the concept of 'commercial communication' implies, as its name suggests, a commercial purpose. For commercial communication, the main rules concern disclosing the commercial character of certain content,[58] and the inclusion of commercial communications in repositories (see section 5 and 6 of this paper). For advertisements, the DSA contains more rules.

### 3.2 The applicability of the DSA to online advertising companies

Online advertising is a complicated business, with many different types of companies, often called adtech companies, involved.[59] Some of these adtech companies can also be considered intermediary services as described in Section 2.2 and are thus covered by the DSA. Other adtech companies are not intermediary services. In this section, we discuss three types of adtech companies. Section 3.2.1 discusses websites and apps with integrated advertising services. These clearly fall in the category intermediary services. Section 3.2.2 shows that advertising networks are also intermediary services. Finally, Section 3.2.3 gives examples of types of adtech companies that are not intermediary services. Therefore, most DSA rules do not apply to this third category.

#### 3.2.1 Online platforms with integrated advertising services

Some online platform companies offer services with integrated advertising. We call the advertising 'integrated' because the company enables advertisers to advertise on the platform directly. Hence, advertisers can advertise on a platform without going through an intermediary. Examples of platforms with integrated advertising services are social network services such as Facebook, Instagram, TikTok and LinkedIn.[60]

---

[57] Article 2(f) of the Directive 2000/31/EC of the European Parliament and of the Council of 8 June 2000 on certain legal aspects of information society services, in particular electronic commerce, in the Internal Market ('Directive on electronic commerce'), OJ L 178, 17.7.2000, p. 1–16, http://data.europa.eu/eli/dir/2000/31/oj.
[58] DSA, art 26(2), 44.
[59] See Michal Veale and Frederik J. Zuiderveen Borgesius, 'Adtech and real-time bidding under European data protection law' (2022) 23 German Law Journal 226.

Such services are hosting services,[61] as they host (store) content for their recipients, who can, for instance, upload pictures and post messages. The services also qualify as 'online platforms' in the sense of the DSA. As noted, an online platform is 'a hosting service that, at the request of a recipient of the service, stores and disseminates information to the public'.[62] On Facebook, for example, recipients can spread (disseminate) pictures and messages to others. Facebook, Instagram, TikTok and LinkedIn are not only online platforms, but also VLOPs, because they have at least 45 million monthly active users.[63] VLOPs must comply with more rules than smaller online platforms.

In this example, the recipients are first and foremost the users of the online platforms that use the platforms to view and disseminate 'organic' content. Additionally, the online platforms are also used by another category of recipients: the advertisers that disseminate their advertisements are also recipients (see Section 2.2).

### 3.2.2. Ad networks

For some types of adtech companies, it is less obvious whether they are covered by the DSA and, if so, how they should be qualified. For example, ad networks are companies that connect advertisers to publishers of apps and websites. In a simplified example, an ad network has contracts with thousands of websites.[64] An advertiser can contract with the ad network to show its ad on those websites.[65] An ad network typically uses tracking cookies or similar technology to recognise individual internet users when they visit websites within the company's network and to build profiles on them.[66] This allows the ad network to present personalised advertisements to the users, for which it can receive a higher remuneration.

---

[60] Recital 13 DSA confirms that social networks sites are online platforms.
[61] DSA, art. 3(g)(iii). See also Section 2.2. of this paper.
[62] DSA, art. 3(i).
[63] See the list of 'Very large online platforms' by the European Commission. https://ec.europa.eu/commission/presscorner/detail/en/ip_23_2413.
[64] In practice, the various activities may be conducted by different parties. See also Section 3.2.3. Furthermore, there may be additional intermediaries between an advertiser and the ad network.
[65] See for an explanation of how ad networks work: Criteo, 'What is an ad network? A guide for advertisers and publishers' (*Criteo.com* 15 September 2023) <https://www.criteo.com/blog/what-is-an-ad-network-a-guide-for-advertisers-and-publishers/> accessed 18 December 2024; Adjust, 'What is an ad network?' <https://www.adjust.com/glossary/ad-network/> accessed 20 December 2024.

As noted, online platforms in the DSA sense (i) are 'hosts' and (ii) disseminate information to a potentially unlimited number of parties.[67] We apply those criteria to our archetype of an ad network.

First, ad networks provide hosting services by storing the advertisements on behalf of the advertisers (the recipients, see also Sections 2.2 and 3.2.1). This was confirmed in 2010, when the Court of Justice of the European Union ruled that one of Google's advertising services ('AdWords') concerned 'hosting' because Google AdWords stored ads for advertisers.[68] That judgment is from more than a decade ago, but it is still relevant because it dealt with a definition of 'host' that closely resembles the one in the DSA.[69]

Second, ad networks also disseminate the stored advertisements to the public. Ad networks disseminate ads without the direct active involvement of the recipient (the provider of the advertisement). Although the recipient may set certain parameters for the dissemination of the advertisement, the ad networks are ultimately responsible for serving the advertisement to specific viewers. In sum, ad networks fall in the category of online platforms.

An ad network may also qualify as a VLOP if it has more than 45 million monthly active recipients in the European Union. Here, the definitions in the DSA are not entirely clear. If an ad network has more than 45 million *advertisers*, the ad network would qualify.

---

[66] For an overview of online tracking: Reuben Binns, 'Tracking on the Web, Mobile and the Internet of Things', *Foundations and Trends in Web Science* 8, no. 1–2 (2022): 1-113, https://arxiv.org/abs/2201.10831, accessed 20 December 2024.

[67] DSA, art. 3(i).

[68] CJEU, Google France SARL and Google Inc. v Louis Vuitton Malletier SA (C-236/08), Google France SARL v Viaticum SA and Luteciel SARL (C-237/08) and Google France SARL v Centre national de recherche en relations humaines (CNRRH) SARL and Others (C-238/08, 22 September 2009, ECLI:EU:C:2010:159, dictum, par. 3.

[69] The eCommerce Directive (2000/31/EC) defines a hosting service in article 14. The DSA defines a hosting service in article 3(g)(iii).

However, it is not immediately clear whether the *viewers* of the advertisements also qualify as recipients of an ad network's service. After all, somebody who visits a website or uses an app does not consciously use, or connect to, an ad network. Many website visitors do not know that ads are provided by other companies than the website. On the other hand, various arguments point to a broader interpretation of the concept of a recipient. First, the definition of recipient does not require that a recipient consciously makes use of a service. Similarly, the definition of an 'active recipient of an online platform' of Article 3(p) DSA only requires that the recipient is 'exposed' to information. According to recital 77, it is not necessary that the recipient actively clicks on the content or performs any other action. The recital also clarifies that an active recipient is not necessarily a registered user.[70] Finally, the viewer in relation to the advertisements is not fundamentally different from a user of an online platform with integrated advertisings services (Section 3.2.1).

Article 3(p) DSA (that defines an 'active recipient of an online platform') requires that the recipient is exposed to information that is 'disseminated through [the platform's] online interface'. However, 'online interface' is defined broadly in Article 3(m) DSA. An online interface is not limited to social media platforms (see Section 3.2.1), but refers to 'any software, including a website *or a part thereof*, and applications, including mobile applications' (emphasis added). Ad networks show advertisements in a designated part of another company's website or app. Such designated parts could be considered a part of the online interface of the ad network.

There is another reason to consider these designated parts as the online interface of an ad network: internet users can interact with an ad. These interactions are distinct from user interactions with other parts of the website. First, a user can usually click on an ad to be redirected by the ad network to the advertiser's website. Second, ad networks such as Google may allow a user to click an icon to get information about the ad or report it.[71] With both interactions, the ad network plays a role that is distinct from the other parts of the website. In sum, viewers of advertisements are exposed to information that is disseminated through the ad network's online interface. We therefore tentatively conclude that such viewers should be considered recipients of the platform service of the ad network.[72]

---

[70] See also 'Questions and Answers on identification and counting of active recipients of the service under the Digital Services Act' (n 25) 4, 6.

It would be useful if regulators clarified whether people who see ads served by ad networks should be counted as 'recipients' of an ad network's services, for example through Article 33(3) DSA. This provision empowers the Commission to adopt delegated acts laying down the methodology for calculating the number of average monthly active recipients. Although this is a slightly different topic than the exact scope of the concept of a recipient, the elaboration of the methodology could require clarifications of the concept as well.[73]

Furthermore, the question of whether the viewers are recipients and the advertisements are shown on the online interface of the ad networks is relevant in the context of the Articles 26 and 28 DSA (on transparency of advertisements and profiling-based advertising targeted at minors). Further clarification can therefore also be based on these articles, for example through guidelines pursuant to Article 28(4), standards pursuant to Article 44(1)(g) and (h), or a code of conduct for online advertising pursuant to Article 46 DSA. In any case, so far, the European Commission has not designated any ad network as a VLOP.[74]

To summarise, we conclude that ad networks serving advertisements on other websites or apps are online platforms and that the viewers of the ads should be considered recipients. We realise that this conclusion may be counterintuitive. At least for the viewers of the ads, the experience of visiting a website and passively being exposed to advertisements is different from the more interactive experience that social media and other typical online platforms with integrated advertising services may offer.

---

[71] See also <https://digitaladvertisingalliance.org/> accessed 17 December 2024; Section 7.1.
[72] Cf Becker and Penfrat (n 7) 58, who state, without providing any arguments, that the DSA does not apply to advertisements served on other websites; Section 7.
[73] Cf 'Questions and Answers on identification and counting of active recipients of the service under the Digital Services Act' (n 25), which also provides limited clarifications.
[74] European Commission, 'Supervision of the designated very large online platforms and search engines under DSA', last updated 2 October 2024, https://digital-strategy.ec.europa.eu/en/policies/list-designated-vlops-and-vloses.

Nevertheless, we feel that this difference should not be overstated. The definition of online platform includes many types of online services. The list of very large online platforms includes social media, online marketplaces, pornography websites, app stores, Google maps, video sharing websites, and an online encyclopaedia (Wikipedia).[75] Considering this wide range of services, it is not far-fetched to see ad networks as platforms. In addition, the potential harms of ads that are disseminated through an ad network resemble the potential harms of ads on social media. There is no clear justification to treat them differently.

### 3.2.3. Adtech companies who do not provide intermediary services

Other adtech companies do not provide intermediary services. For example, a company that specialises in analysing the number of people that ads reach may not host any information for others. If a company does not host information for others, it does not provide hosting or platform services. Similarly, an ad verification company that checks whether other companies serve ads in a way that internet users can actually see them may not host any advertisements either.[76] Hence, some adtech companies are not intermediary services in the sense of the DSA. Therefore, most DSA rules do not apply to them.

Adtech companies can come in many variations. A complicated economy of companies has developed around online advertising. For instance, many ads are shown on websites after a 'real time bidding' process. Real time bidding is a system where pre-determined advertising space, such as a banner advert on a website, or a splash screen in an app, is allocated through an auction process, which is carried out each time an ad is shown.[77]

---

[75] https://digital-strategy.ec.europa.eu/en/policies/list-designated-vlops-and-vloses
[76] To illustrate: a company called DoubleVerify offers such a service. <https://doubleverify.com/viewability/> accessed 20 December 2024.
[77] See M. Veale & F.J. Zuiderveen Borgesius, 'Adtech and real-time bidding under European data protection law', German Law Journal, 23(2), 226-256, 2022.

Billions of such automated auctions take place every day. Different types of companies can be involved in real time bidding and online advertising, such as 'supply side platforms', 'demand side platforms', and 'ad exchanges'.[78] A discussion of whether each type of adtech company would qualify as an 'online platform' falls outside the scope of this paper. There may be much debate, and perhaps case law, about whether certain types of adtech companies count as online platforms or not. Guidance by regulators on this point would be welcome.

## 4    The application of the general rules of the DSA to online advertising

The previous Section shows that the general concepts of the DSA also apply to online advertising. Furthermore, the advertisements themselves are disseminated through online platforms. Advertising is information and can constitute illegal content. For this reason, the general rules of the DSA also apply to online advertisements. Although a discussion of all the relevant rules of the DSA is outside the scope of this paper, a short description of the application of the general rules is needed to understand the implications of the DSA for online advertising.

This Section will follow the layered approach of the DSA.[79] The Section starts with the general rules for all online intermediary services and subsequently discusses the additional obligations for hosting services, online platforms, online marketplaces and VLOPs and VLOSEs.

### 4.1  Obligations for all providers of intermediary services

All providers of intermediary services must be transparent about their content moderation. First, Article 14(1) DSA concerns content moderation *policy*. The providers should include information about any restrictions that they impose in relation to the use of their service in their terms and conditions. This includes information on their content moderation policies, procedures, measures and tools used. In relation to advertisements, this means that the providers should be clear about what kind of advertising is allowed or not and about how they assess the permissibility. The information should be in clear language and easily accessible. The providers should apply these restrictions in a 'diligent, objective and proportionate manner'.[80]

---

[78] See idem.
[79] See Section 2.2 of this paper about the DSA's layered approach.

Article 15 DSA imposes an obligation on providers to publish publicly available reports on the content moderation *that they engaged in*. This obligation and the other general reporting obligations are further discussed in conjunction with the more specific reporting obligations for online advertisements (Section 6).

### 4.2 Obligations for providers of hosting services

The providers of hosting services, including online platforms, must put in place a notice and action mechanism.[81] Such a system allows anyone to notify the provider that certain content is illegal. It should be easy to access and user-friendly. As advertisements can also be illegal content, the providers should also facilitate the notification of illegal advertisements. The providers should confirm the receipt of the notice, process the notices they receive and inform the notifier about their decision in respect of the notified advertisement.[82]

If the provider removes the advertisement or otherwise restricts its visibility, it should give a statement of reasons to the provider of this advertisement.[83] In contrast, the DSA does not oblige the provider to explain the decision to not remove the advertisement to the notifier.[84] If the advertisement is illegal but the provider does not remove it after a notification, the exemption from liability no longer applies.[85]

---

[80] DSA, art 14(4).
[81] DSA, art 16.
[82] DSA, art 16(4), (5), (6).
[83] DSA, art 17(1)(a).
[84] See also Wolters and Gellert (n 11) 415.
[85] DSA, art 6(1), 16(3). See also Section 2.1.

### 4.3 Obligations for online platforms

The DSA imposes several redress mechanisms in relation to the content moderation decisions of online platforms. If the notifier or the party who submitted the advertisement disagree with a decision to remove or not remove the advertisement or otherwise restrict its visibility, they can contest this decision through an internal complaint-handling system [86] and through out-of-court dispute settlement. [87] If a recipient frequently provides manifestly illegal advertisements, the online provider should suspend the provision of their services to these recipients.[88] Similarly, platforms should not process notices and complaints from individuals that frequently submit notices or complaints that are manifestly unfounded.[89]

Online platforms should also facilitate notifications by 'trusted flaggers' acting in their designated area of expertise and to process them without undue delay.[90] The status of the trusted flaggers shall be awarded by the national Digital Services Coordinators.[91] As industry organisations can also be trusted flaggers,[92] it is likely that advertisements will also be relevant to many of these flaggers. For example, an industry organisation may want to report advertisements for counterfeit goods or services that violate copyright law. Fact-checking organisations could notify political advertisements containing (illegal)[93] disinformation.

---

[86] DSA, art 20.
[87] DSA, art 21.
[88] DSA, art 23(1).
[89] DSA, art 23(2).
[90] DSA, art 22(1).
[91] DSA, art 22(2).
[92] DSA, recital 61.
[93] The obligation to in place a notice and action mechanism and facilitate notifications by trusted flaggers is limited to notifications about illegal content. DSA, art 16(1), recital 61. However, providers may decide to also facilitate notifications of permissible content that contains disinformation or otherwise violates the terms and conditions of the platform. This is especially relevant for VLOPs and VLOSEs (see also Section 4.4) or for platforms that committed to the 'Strengthened Code of Practice on Disinformation 2022'. *The Strengthened Code of Practice on Disinformation 2022* <https://digital-strategy.ec.europa.eu/en/library/2022-strengthened-code-practice-disinformation> accessed 20 December 2024, commitments 30, 31, 32; Gellert and Wolters (n 14) 46.

## 4.4 Obligations for online marketplaces

The DSA obligates online marketplaces to ensure that traders[94] can only use the service if they first supply information that ensures their traceability.[95] The online marketplaces also have some obligations to verify this information [96] and to inform affected consumers if they become aware that an offered product or service is illegal.[97] Finally, online marketplaces should design their interface in a way that enables traders to comply with their information obligations.[98]

The obligation to ensure traceability applies if the traders use the online marketplaces to offer products or services, but also if they use the platform to promote messages on these products or services. However, the obligation only applies to platforms that can be used to actually conclude the contract. In this light, the obligation does not apply to online advertisements on platforms that redirect the users to another website that can be used to purchase the product or service.

## 4.5 Obligations for very large online platforms and search engines

Very large online platforms (VLOPs) and very large online search engines (VLOSEs) are obligated to diligently identify, analyse and assess any systemic risks stemming from the designing or function of their service[99] and take measures to mitigate these risks. [100] The European Commission may also order VLOPs and VLOSEs to take measures in the event of a crisis in relation to public security or public health.[101] Finally, VLOPs and VLOSEs must subject themselves to an independent audit in relation to their content moderation obligations once a year and take the necessary measures to implement the operational recommendations if the audit report is not positive.[102]

---

[94] Traders are defined in DSA, art 3(f).
[95] DSA, art 30.
[96] DSA, art 30(2), (3). See also DSA, art 31(3).
[97] DSA, art 32.
[98] DSA, art 31.
[99] DSA, art 34.
[100] DSA, art 35.
[101] DSA, art 36(1), (2).
[102] DSA, art 37(1), (6).

Advertisements may also lead to systemic risks. For example, ads may be illegal or contain disinformation that can have negative effects on civil discourse, electoral processes, public security or public health.[103] If this is the case, the providers can be obligated to take measures in relation to their advertising systems, recommender systems, or content moderation policy and processes.[104] Providers of VLOPs and VLOSEs may also need to put in place measures if the content moderation of the advertisements negatively affects fundamental rights, such as the right to freedom of expression.[105]

## 5    Transparency requirements

The DSA imposes several transparency obligations in relation to advertisements. First, online platforms should be transparent about the functioning of their recommender systems (Section 5.1). Furthermore, they should provide various information about the nature and origin of the presented advertisements (Section 5.2).

### 5.1 Transparency of recommender systems

The DSA contains various rules in relation to recommender systems. This Section analyses the transparency requirements in relation to recommender systems. We discuss the restrictions on profiling in Section 7.1. The DSA defines 'recommender system' as a system to suggest or prioritise information.[106] Since advertisements are also information (see Section 3), this definition also applies to systems that suggest or prioritise advertisements.

Article 27 DSA obligates online platforms to be transparent about its recommender systems. Online platforms should set out the main parameters and their relative importance in the terms and conditions. If the recipients can modify or influence the main parameters, this functionality should be directly and easily accessible.[107]

---

[103] DSA, art 34(1)(a), (c), (d).
[104] DSA, art 35(1)(b), (c), (d), (e).
[105] DSA, art. 34(1)(b). See also Section 2.1.
[106] Article 3(s)

Article 27 DSA obligates the online platforms to be transparent about the *general* functioning of their recommender systems for all content. Article 26(1)(d) provides a more stringent rule for advertisements. Online platforms should provide meaningful information about the main parameters used to determine the recipient to whom the advertisement is presented. This information should be presented for each *specific* advertisement presented to each individual recipient. The information should be directly and easily accessible from the advertisement.

Article 26(1)(d) DSA thus obligates the online platforms to be transparent about *individual* advertisements. At the same time, the provision does not clarify how individualised the information should be. For example, assume that an ad is targeted to males under 30. Should the online platform specifically clarify this, or is it sufficient to specify that the advertisement is presented based on the gender and age of the recipient? The requirement that the information should be 'meaningful' suggests the former. However, the required level of detail remains unclear.[108] The DSA stipulates that voluntary standards can give more details about how to organise (the technical aspects of) transparency for advertising.[109]

---

[107] DSA, art 27(3).
[108] A short personal experiment on 15 November 2024 shows that VLOPs offer different levels of details about ads. For example, Facebook has presented one of the authors an advertisement based on the fact that he communicated in English, set his age to 25 and older, has a primary location in the Netherlands and has previously interacted with ads about 'video games, productivity and clothing' and pages and posts about 'music, travel and graphic novels'. Facebook also provided a link to general information about how other factors may be relevant. Presumably, these secondary factors are thus not the main parameters referred to in Article 26(1)(d) DSA. LinkedIn is a bit less clear. It states that an advertisement is shown because of, for example, the size and 'growth rate' of the current employer and the location and inferred job seniority of the user without explicitly explaining *how* these parameters are relevant. For example, several advertisements were shown to the author because of his country. However, LinkedIn did not clarify whether the advertisement was specifically targeted to this country or whether the country was part of a broader targeted area.
[109] DSA, art 44(1)(g), (h), 46.

Ad networks typically use systems that suggest or prioritise ads and thus typically use recommender systems. Since ad networks must be seen as online platforms (see Section 3.2.2), they must comply with the rules on recommender systems.

## 5.2 Additional transparency requirements for individual advertisements

Article 26 DSA contains more requirements regarding transparency about individual advertisements. Platforms should always identify an advertisement as such. Online platforms should ensure that recipients are always able to identify that the information is an advertisement, the person on whose behalf it is presented, and the person who paid for it.[110]

The DSA also contains a similar rule for other kinds of commercial communications. Pursuant to Article 26(2) DSA, online platforms should allow recipients to declare that the content they provide is or contains commercial communications. If a recipient uses this functionality, the online platforms should make sure that other recipients can always identify the commercial communications as such.

Article 26(2) DSA only applies when the recipients themselves declare that the content contains commercial communications. Online platforms are not obligated to proactively or reactively monitor whether this is the case.[111] In this light, the obligation is designed to complement and facilitate the compliance with other obligations to disclose the commercial nature of communications.[112]

---

[110] DSA, art 26(1)(a), (b), (c). See also Section 6.2.
[111] Cf DSA, art 8. A reactive obligation does exist if the commercial communication is illegal content, for example because it is misleading about its commercial intent. Cf DSA, art 16; n 111.
[112] DSA, recital 68, referring to Directive 2010/13/EU of the European Parliament and of the Council of 10 March 2010 on the coordination of certain provisions laid down by law, regulation or administrative action in Member States concerning the provision of audiovisual media services (Audiovisual Media Services Directive) [2010] OJ L95/1, as amended by Directive (EU) 2018/1808 of the European Parliament and of the Council of 14 November 2018 amending Directive 2010/13/EU on the coordination of certain provisions laid down by law, regulation or administrative action in Member States concerning the provision of audiovisual media services (Audiovisual Media Services Directive) in view of changing market realities [2018] OJ L303/69, art 9; Directive 2005/29/EC of the European Parliament and of the Council of 26 May 2005 concerning unfair business-to-consumer commercial practices in the internal market and amending Council Directive 84/450/EEC, Directives 97/7/EC, 98/27/EC and 2002/65/EC of the European Parliament and of the Council and Regulation (EC) No 2006/2004 of the

In contrast, online platforms will always know when they receive remuneration for promoting certain content and the content thus qualifies as an advertisement (see also Section 3). For this reason, the obligation to be transparent about advertisements does not depend on a declaration of the recipient that provides the advertisement.

Disclosure obligations like in Article 26 DSA are not new. For example, the eCommerce directive from 2000 already required that commercial communication is clearly identifiable as such.[113] The Strengthened Code of Practice on Disinformation 2022 also includes disclosure obligations.[114] However, this code is limited to political or issue advertisements. All in all, the DSA continues the tradition in EU law that advertisements should be clearly identifiable as such.

## 6    Reporting obligations and advertising repositories

In addition to the transparency requirements for individual advertisements (Section 5.2), the DSA also imposes various reporting requirements. It contains both general reporting obligations that also apply to online advertisements (Section 6.1) and a more specific obligation to compile an advertisement repository (Section 6.2).

These obligations are of a different nature compared to those discussed in the previous Section. Whereas the transparency requirements for individual advertisements are directed towards the recipients that view them, the reporting obligations discussed in this Section are directed towards the general public. The reporting obligations also allow supervisory authorities to hold the providers of the intermediary services accountable.[115]

---

European Parliament and of the Council ('Unfair Commercial Practices Directive') [2005] OJ L149/22, art 7(2).
[113] Article 6(a), eCommerce Directive 2000/31/EC.
[114] Strengthened Code of Practice on Disinformation 2022 (n 92), Commitments 6, 8.
[115] See DSA, recital 49.

## 6.1 General reporting obligations

The DSA contains several general reporting obligations for online intermediaries. These obligations are mostly related to content moderation. Since advertisements are also a form of content (see also Section 3), these general obligations are also relevant in relation to online advertising. Again, the obligations follow the layered approach of the DSA.

First, pursuant to Article 15 DSA, all providers of intermediary services must publish yearly reports about any content moderation that they engaged in.[116] The requirements on these reports are fairly detailed. For example, the providers should report the number of received orders from authorities, the number of complaints through the internal complaint-handling system and the number and types of the taken content moderation measures.[117] Hosting services should also report the number of received notices.[118] Finally, pursuant to Article 24 DSA, online platforms must report the number of disputes submitted to the out-of-court dispute settlement bodies and the number of suspensions imposed for the provision of manifestly illegal content, unfounded notices and unfounded complaints.[119]

All of these obligations also apply to orders, complaints, measures, notices, disputes or suspensions in relation to online advertisements. However, the Articles 15 and 24 DSA do not require the providers to differentiate between advertisements and other content. Article 15 DSA merely states that the information should be categorised by the 'type of illegal content'. Although not entirely clear, various recitals suggest that the term

---

[116] DSA, art 15
[117] DSA, art 15(1)(a), (c), (d). In relation to the internal complaint-handling system, art 15(1)(d) differentiates between complaints in accordance with the terms and conditions (for all providers) and in accordance with Article 20 (only for online platforms).
[118] DSA, art 15(1)(b). Interestingly, the hosting services should also specify the number of notices submitted by trusted flaggers, even though only online platforms are obligated to facilitate notifications by trusted flaggers. Cf the more nuanced formulation of art 15(1)(d); n 116. See also DSA, art 16(3). Trusted flaggers are obligated to publish yearly reports on the notices submitted by them.
[119] DSA, art 24(1).

'type of illegal content' refers to the *reason* for the illegality[120] and not to the type of content (e.g. advertisement, picture, video, text).

Articles 15 and 24 DSA thus only provide limited transparency in relation to online advertising specifically. However, the DSA empowers the European Commission to adopt implementing acts to lay down templates concerning the form, content and other details of the reports.[121] Through this mechanism, the European Commission could create more transparency by requiring providers to differentiate between advertisements and other types of content. The European Commission has not yet adopted such an implementing act.

VLOPs and VLOSEs have additional reporting obligations.[122] They should report twice a year,[123] provide more details about the invested human resources[124] and report on the fulfilment of the additional obligations for VLOPs and VLOSEs.[125] Again, these obligations do not require separate information in relation to advertisements.

Finally, the DSA obligates VLOPs and VLOSEs to grant access to data.[126] The access exists for two groups. First, the access can be requested by the European Commission or the 'Digital Services Coordinator of establishment'[127] if the data are necessary for effective supervision.[128]

---

[120] DSA, recital 17 ('any type of illegal content, irrespective of the precise subject matter or nature of those laws'), 52 ('in particular by taking into account the type of illegal content being notified and the urgency of taking action. For instance, such providers can be expected to act without delay when allegedly illegal content involving a threat to life or safety of persons is being notified'). See also DSA, art 15(1)(c). Information in relation to the content moderation engaged in at the providers' own initiative should be categorised by the type of illegal content, but also by the violation of the terms and conditions. Again, the reason for the content moderation ('violation') determines the categorisation.
[121] DSA, art 15(3) and 24(6).
[122] DSA, art 42.
[123] DSA, art 42(1).
[124] DSA, art 42(2).
[125] DSA, art 42(4).
[126] DSA, art 40.
[127] This is, in short, the Digital Services Coordinator of the member state where the main establishment or legal representative of the provider is located. DSA, art 3(n).
[128] DSA, art 40(1), (2).

Second, the Digital Service Coordinator of establishment can also request access for 'vetted researchers' for the purpose of research in relation to systemic risks.[129] Since online advertisements can also contribute to systemic risks (Section 4.5), such researchers can also request data in relation to advertisements. VLOPs and VLOSEs can refuse the request if they do not have the data.[130] Article 40 thus does not impose additional obligations to collect or store data in relation to advertisements.

## 6.2 Advertising repositories

Article 39 DSA imposes an additional specific transparency obligation in relation to advertisements. VLOPs and VLOSEs that present advertisements should compile a publicly available repository with information about the presented advertisements. The repository should facilitate supervision and research into risks related to advertisements.[131] It should therefore be publicly available through a searchable and reliable tool that allows multicriteria queries and through application programming interfaces.[132]

The repository should contain information on the advertisements that are currently presented on the platform, but also on advertisements that have been visible in the last year.[133] This is an improvement over existing repositories.[134] DSA repositories do not only provide a more complete picture, they also create transparency about the content moderation of the advertisements. Furthermore, the repository should also contain other

---

[129] DSA, art 40(4), (8). About this access, see e.g. Laura Edelson, Inge Graef and Filippo Lancieri, *Access to Data and Algorithms: For an Effective DMA and DSA Implementation* (CERRE 2023) 54-72.
[130] DSA, art 40(5)(a).
[131] DSA, recital 95.
[132] DSA, art 39(1).
[133] DSA, art 39(1).
[134] Meta added this functionality one week before this became an obligation under the DSA. See <https://www.facebook.com/ads/library/> accessed 20 December 2024. Under 'See what's new', the website states: 'As of 17 August 2023, the Meta Ad Library displays and archives ads that deliver an impression to the EU and its associated territories, along with the ads' targeting and delivery information. You can also search for information about branded content on Facebook and Instagram.' It already existed for political advertisements. Cf n 138. About this library, see e.g. Paddy Leerssen and others, 'News from the ad archive: how journalists use the Facebook Ad Library to hold online advertising accountable' (2023) 26 Information, Communication & Society 1381, 1381-1383.

commercial communications that were identified as such by the recipients that provided them.[135]

Article 39(2) DSA stipulates the necessary details of the repository. It should contain information about the advertisement itself, such as its content (a), on whose behalf it is shown (b), who paid for it (c) and when it was shown (d).[136] Furthermore, the repository should contain information about the recipients of the advertisement. The repository should specify whether the advertisement was targeted to one or more particular groups of recipients and the main parameters used for this purpose (e) and the total number of recipients that viewed the advertisement, further broken down per member state or targeted group (g). The European Commission can issue guidelines to further specify the requirements on the structure, organisation and functionalities of the repositories.[137]

If VLOPs or VLOSEs removed an advertisement because it contains illegal content or violated their terms and conditions, the repository should not show the above-mentioned information specified in Article 39(2)(a) through (c). Instead, the repository should specify the reasons for the removal or, if removed after an order from a competent authority, the legal basis for this order.[138] Although this rule prevents the further dissemination of removed illegal content through the repository, it also reduces the transparency about the content moderation of advertisements.

For example, it becomes harder to analyse how the platform moderates different kinds of disinformation disseminated through advertisements if it no longer possible to compare the content of various advertisements. In this light, a more fine-grained solution would have been preferable. Whereas advertisements that are illegal in any context (e.g. terrorist content, child sexual abuse material, advertisements for weapons or counterfeit products) should not be visible in the repository, this is not necessary in the case of harmful content that only leads to risks under certain circumstances (e.g. disinformation).

---

[135] DSA, art 39(2)(f).
[136] The information under (b) and (c) should also be presented to the recipients that see the advertisements. Cf DSA, art 26(1)(b), (c).
[137] DSA, art 39(3).
[138] DSA, art 39(3).

The obligation to maintain an advertisement repository is not entirely new. A comparable obligation is also included in the Strengthened Code of Practice on Disinformation 2022. [139] However, this code is limited to political or issue advertisements. The DSA extends the obligation to all advertisements and provides a more stringent legal basis.

## 7    Restrictions on profiling

The DSA contains various restrictions on 'profiling as defined in Article 4, point (4), of [the GDPR]'. This Section describes the restrictions on the use of profiling for recommender systems (Section 7.1) and advertisements (Sections 7.2 and 7.3) in turn.

Article 4(4) of the GDPR defines profiling as 'any form of automated processing of personal data consisting of the use of personal data to evaluate certain personal aspects relating to a natural person, in particular to analyse or predict aspects concerning that natural person's performance at work, economic situation, health, personal preferences, interests, reliability, behaviour, location or movements'.

Behavioural advertising is an example of a practice that falls within the GDPR's profiling definition. Behavioural advertising is a type of targeted online advertising that involves monitoring people's online behaviour and using the collected information to show people targeted advertisements. [140] In a simplified behavioural advertising scenario, an ad network company automatically analyses somebody's web browsing behaviour (for instance with a tracking cookie), to predict that person's interests, and the chance that the person will click on ads about certain topics. For instance, a company may conclude that somebody who visits many websites about recipes likes cooking and may therefore click on ads for cookbooks.

---

[139] Strengthened Code of Practice on Disinformation 2022 (n 92), Commitments 10, 11. The details are similar, see e.g. Measure 10.1. However, the information should be available for the longer period of 5 years under the Strengthened Code of Practice on Disinformation. Strengthened Code of Practice on Disinformation 2022 (n 92), Measure 10.2.
[140] Boerman, Kruikemeier and Zuiderveen Borgesius (n 2)363-376; Aleksandre Zardiashvili (Lex Zard), *Power and dignity: the ends of online behavioral advertising in the European Union* (PhD thesis University of Leiden 2024) <https://hdl.handle.net/1887/3753619> accessed 20 December 2024.

## 7.1 Opt-out of profiling for recommender systems

VLOPs and VLOSEs have additional obligations regarding recommender systems. In addition to the transparency requirements discussed in Section 5.1, VLOPs and VLOSEs must always offer one option that is not based on profiling pursuant to Article 38 DSA. Hence, roughly speaking, VLOPs and VLOSEs must offer a version of the recommender system that is not personalised. Since recommender systems also apply to advertisements (see also Section 5.1), this means that recipients of VLOPs and VLOSEs should always have the option to turn off personalisation of advertisements that is based on profiling. If an ad network company is so large that it counts as a VLOP, it should also comply with Article 38 DSA and offer a profiling-free version of its recommender system.

Ad networks sometimes enable internet users to stop personalised ads. For example, Google enables advertisers to advertise on websites through Google's ad network. Google enables a visitor to a non-Google website (e.g. cnn.com) to click on a small logo, for instance a small 'triangle', in an ad that is served by Google. After clicking that logo, the website visitor sees a Google web page where the visitor can choose to 'Turn off Personalize [sic] ads on this site.'[141] Criteo, another ad network company, offers a similar option.[142] But the fact that Google and Criteo offer this possibility does not necessarily mean that the companies see their ad networks as VLOPs. Online marketing companies have offered some possibilities to opt out of certain forms of profiling-based advertising for years – long before the application of the DSA.[143]

---

## 7.2 A ban on profiling-based advertising using sensitive data

During the negotiations about the draft DSA, some members of the European Parliament wanted to ban all profiling-based advertising (behavioural advertising).[144] Such a ban did not make it to the final text of the DSA. But the DSA does include two bans: on profiling-based advertising based on sensitive data, and on profiling-based advertising targeted at children. This and the next Section discuss these bans.

We start with the ban on profiling-based advertising using 'special categories of personal data', often called sensitive data in practice.[145] The DSA's preamble suggests that the ban aims, among other things to mitigate the risk of manipulative or discriminatory advertising.[146] Article 26(3) DSA says:

> Providers of online platforms shall not present advertisements to recipients of the service based on profiling as defined in Article 4, point (4), of [the GDPR] using special categories of personal data referred to in Article 9(1) of [the GDPR].

The GDPR contains extra strict rules on 'special categories of personal data' and defines those as 'personal data revealing racial or ethnic origin, political opinions, religious or philosophical beliefs, or trade union membership, and (…) genetic data, biometric data for the purpose of uniquely identifying a natural person, data concerning health or data concerning a natural person's sex life or sexual orientation'.[147]

---

[144] See the 'Tracking-Free Ads Coalition', https://trackingfreeads.eu/coalition-members/ accessed 20 December 2024. See also 'How corporate lobbying undermined the EU's push to ban surveillance ads' (*Corporate Europe Observatory* 18 January 2022 <https://corporateeurope.org/en/2022/01/how-corporate-lobbying-undermined-eus-push-ban-surveillance-ads> accessed 20 December 2024; Clothilde Goujard, 'European Parliament pushes to ban targeted ads based on health, religion or sexual orientation' (*Politico* 20 January 2022) <https://www.politico.eu/article/european-parliament-bans-use-of-sensitive-personal-data-for-targeted-ads/> accessed 20 December 2024.

[145] Calling 'special categories' of data 'sensitive data' is not fully correct. All types of personal data can be more or less sensitive, depending on the context. For instance, one's home address is not sensitive for some people. But victims of a stalker might want to keep their house address secret; for such victims their home address is sensitive.

[146] DSA, recital 69. See generally on this ban Zardiashvili (n 139) section 6.2.1.

[147] GDPR, art 9(1).

### Scope of the ban: profiling

The ban only applies to advertising 'based on profiling'. Targeted advertising that is not based on profiling is thus still allowed.[148] As noted, 'profiling' in the sense of the GDPR requires that personal data are processed in an automated way 'to *evaluate* certain personal aspects relating to a natural person'.[149]

For example, if somebody declares that he or she is gay, and a marketing company uses that information to target ads to that person, the company has arguably not 'evaluated' aspects of the person. Therefore, the company has not engaged in profiling, and the restrictions on profiling do not apply.[150]

Contextual advertising is also outside the scope of profiling. Contextual advertisements are not personalised based on the characteristics of individual viewers. Instead, they are tailored to the content next to which they are shown. For example, ads for tennis rackets may be placed on a website about tennis. Similarly, ads for weight loss pills may be placed next to content about obesity or eating disorders. Although such contextual advertising can have adverse effects, they do not involve processing personal data. They thus fall outside the scope of the restrictions on profiling.

The restrictions of the Articles 26 and 28 are limited to 'their online interfaces', 'recipients of the service' or 'their interface'.[151] This raises the question whether the restrictions cover ad networks that present advertisements on third party websites. In this situation, the advertisements are arguably shown on the interface and to the recipients of a third party.[152] However, in Section 3.2.2 we showed that this interpretation is probably incorrect, and that a part of the third-party website or app may be considered the interface of the ad network.

---

[148] Duivenvoorde and Goanta (n 7) 9.
[149] GDPR, art 4(4) GDPR, emphasis by the authors. See also Article 29 Data Protection Working Party, *Guidelines on Automated individual decision-making and Profiling for the purposes of Regulation 2016/679* (17/EN WP251rev.01, 2018) 7.
[150] Such personalised advertising is still subject to other restrictions, such as the GDPR's rules on special categories of personal data.
[151] DSA, art 26(1), (3), 28(2).
[152] See Becker and Penfrat (n 7) 58.

### Scope of the ban: 'special categories of personal data'

As mentioned, the ban applies to profiling-based advertising with 'special categories of personal data' as defined in Article 9 GDPR.[153] Case law of the CJEU shows that the concept of 'special categories of personal data' must be interpreted broadly. For example, in 2022, the CJEU said, in summary, that publishing the names of two people online and adding that they are in a relationship reveals their sexual orientation if the names disclose the gender of the two people.[154] In 2023, the CJEU said in the Meta/Bundeskartellamt judgment that, in short, collecting data on somebody's web surfing behaviour or app use entails the processing of special category data if those sites or apps relate to a special category of data.[155] In 2024, the CJEU confirmed in the Lindenapotheke judgment that data concerning health 'must be interpreted broadly'.[156] The CJEU added that if an online store registers that somebody orders a medicinal product at the store, that store processes personal data concerning health, a type of special categories of data.[157] In sum, the concept of 'special categories of personal data' must be interpreted broadly, also in the context of profiling-based advertising. Therefore, many types of profiling-based advertising may fall within the scope of this ban.

### Scope of the ban: 'using' special categories of personal data

The ban of Article 26(3) DSA only applies to profiling-based advertising 'using special categories of personal data'. How should the word 'using' be interpreted?

---

[153] Article 26(3) DSA.

[154] CJEU, case Case C-184/20, ECLI:EU:C:2022:601, OT v. Vyriausioji tarnybinės etikos komisija, 1 August 2022. The CJEU said that the use 'of the verb "reveal" is consistent with the taking into account of processing not only of inherently sensitive data, but also of data revealing information of that nature indirectly (…)', par. 123. See also: Hoppe, T. 'Anticorruption, Privacy, and (Gay) Family–The CJEU Rules on Financial Disclosure of Public Officials (C-184/20);. *ICL Journal, 17*(4), 2023, 433-446.

[155] CJEU, Case C-252/21, ECLI:EU:C:2023:537, 4 July 2023, Meta/Bundeskartellamt, par. 73. In a different case, a Dutch court arrived at a similar conclusion: Rechtbank Amsterdam, Data Privacy Stichting v. Meta, 15 March 2024, ECLI:NL:RBAMS:2023:1407 https://deeplink.rechtspraak.nl/uitspraak?id=ECLI:NL:RBAMS:2023:1407 accessed 20 December 2024.

[156] CJEU 2024, Case C-21/23, ECLI:EU:C:2024:846, Lindenapotheke, 4 October 2024, par 81.

[157] CJEU 2024, Case C-21/23, ECLI:EU:C:2024:846, Lindenapotheke, 4 October 2024.

Suppose that a social network company offers profiling-based advertising services: advertisers can target ads to people, based on categories of people that the social network has defined. The social network company assigns people to categories based on their online behaviour. Suppose that the company registers that somebody often visits a page (on the social network) about kosher recipes. The company registers these visits. Arguably, those visits show that the person is probably Jewish. In the light of the CJEU's Meta/Bundeskartellamt judgment,[158] the company is 'using special categories of personal data'.[159] After all, our hypothetical company registers visits to pages that can show, or at least suggest, somebody's religion.

The company might even fall within the scope of the ban if it does not enable advertisers to target people *based* on special categories of data. For instance, suppose that the company enables advertisers to target people who are interested in 'recipes'. In that case, the *targeting* does not involve special category data. But the company is still 'using' such special category data. After all, the company registers visits that suggest that somebody is probably Jewish. This broad interpretation of the ban seems to be closest to the wording of the ban.

A narrower interpretation of the ban might also be possible. In a narrower interpretation, the ban only applies if a platform enables advertisers to *target* people based on special category data. Under this narrow interpretation, the ban does not apply if the hypothetical social network allows advertisers to target a user because of his or her interest in 'recipes'. But if the social network enables advertisers to target this person because he or she is in the category 'Jewish', the ban applies.

However, the broader interpretation of the ban is more convincing, because the broader interpretation is closer to the text of Article 26(3) DSA. Moreover, recital 69 DSA confirms that the broader interpretation is appropriate. The recital says: 'providers of online platforms should not present advertisements based on profiling as defined in [the GDPR], using special categories of personal data referred to in Article 9(1) of that

---

[158] CJEU, Case C-252/21, ECLI:EU:C:2023:537, 4 July 2023, Meta/Bundeskartellamt.
[159] Article 26(3) DSA.

Regulation, *including* by using profiling categories based on those special categories' (emphasis added). Hence, the recital suggests that the ban also applies if a company is using special category data for profiling-based advertising but does not enable advertisers to target ads based on those special category data. After all, the recital says that the ban applies to using special category data, 'including' when the company enables advertisers to target based on those special category data. If the ban were limited to such targeting based on special category data, the word 'including' in the recital would make no sense.

In conclusion, the DSA's ban of using special categories of personal data for profiling-based advertising has a rather broad scope. But there is still room for discussion about the exact scope of the ban. It would be useful if regulators clarified how the ban should be interpreted, for example in conjunction with the development and implementation of voluntary standards under Article 44(1)(h) DSA.

### What does Article 26(3) DSA add to existing law?

Existing law already places requirements on profiling-based advertising, especially when special categories of data are used. The DSA does not affect ('is without prejudice to') the rules of the GDPR and the ePrivacy Directive.[160] Data protection law may thus impose stricter requirements. Furthermore, data protection law still applies to advertising practices that fall outside the scope of Article 26(3) DSA.

The GDPR and the ePrivacy Directive already require 'explicit consent' of the internet user for profiling-based targeting based on special categories of data.[161] The GDPR applies when 'personal data' are 'processed'. Personal data are defined broadly in the GDPR; in short, any information relating to an identified or identifiable natural person ('data subject') is personal data. Tracking cookies, device fingerprints, and similar online identifiers are personal data too.[162] Processing is defined even more broadly:

---

[160] DSA, art 2(4)(g).
[161] ePrivacy Directive, art 5(3).
[162] GDPR, art 4(1) and recital 30. See Frederik J. Zuiderveen Borgesius, 'Singling out people without knowing their names – behavioural targeting, pseudonymous data, and the new Data Protection Regulation' (2016) 32 Computer Law & Security Review 256.

virtually everything that can be done with personal data counts as processing.[163] Hence, online profiling-based advertising generally involves processing personal data, and thus the GDPR applies.

As noted, the GDPR has extra strict rules on 'special categories of data'. The use of such data 'shall be prohibited', says Article 9 GDPR.[164] The GDPR gives exceptions to that ban. For profiling-based targeting, the only possible exception is the 'explicit consent' of the internet user ('data subject' in GDPR parlance).[165]

Apart from that, the GDPR only allows personal data processing if the relevant company has a 'legal basis' for that processing.[166] In principle, there are six possible legal bases.[167] However, the European Data Protection Board says, roughly summarised, that the internet user's consent is the only possible legal basis for most types of online targeted advertising.[168] Many scholars agree.[169] The DSA's preamble confirms that the GDPR requires that companies obtain consent before they conduct profiling-based targeting.[170] In addition, the ePrivacy Directive requires companies to obtain the internet user's consent for the use for tracking cookies or similar tracking technologies.[171] The requirements for valid consent are strict.[172]

---

[163] GDPR, art 4(2).
[164] GDPR, art 9(1).
[165] GDPR, art 9(2)(a).
[166] GDPR, art e 6(1). For ease of reading we speak of 'companies'. The GDPR actually refers to 'controllers'; see GDPR, art 4(7).
[167] GDPR, art 6(1).
[168] European Data Protection Board, 'Binding Decision 4/2022 on the dispute submitted by the Irish SA on Meta Platforms Ireland Limited and its Instagram service (Art. 65 GDPR)' (5 December (2022) <https://edpb.europa.eu/our-work-tools/our-documents/binding-decision-board-art-65/binding-decision-42022-dispute-submitted_en> accessed 20 December 2024, in particular par. 421.
[169] See e.g. Jiahong Chen, *Regulating Online Behavioural Advertising Through Data Protection Law*. (Edward Elgar Publishing 2021), chapter 5; Zardiashvili (n 139) 214; Frederik J. Zuiderveen Borgesius, 'Personal data processing for behavioural targeting: which legal basis?', International Data Privacy Law 2015-5-3, p. 163-176, 2015
[170] DSA, recital 68: 'Recital 68 says that the DSA 'is without prejudice to the application of the relevant provisions of Regulation (EU) 2016/679 [the GDPR], (…) specifically the need to obtain consent of the data subject prior to the processing of personal data for targeted advertising.'
[171] ePrivacy Directive, art 5(3).
[172] Valid consent requires an 'unambiguous indication of the data subject's wishes by which he or she, by a statement or by a clear affirmative action, signifies agreement to the processing of personal data relating to him or her' (art 4(11) GDPR). Moreover, valid consent requires that consent is 'informed' and 'freely given', in other words: voluntary (art 4(11) GDPR).

In sum, EU law already requires prior explicit consent for profiling-based advertising using special categories of data. The DSA's ban on profiling-based advertising and sensitive data is stricter than the existing rules, because the DSA does not enable people to override the ban with consent.

Facebook (Meta) and Google (Alphabet) decided in 2021 to stop the possibility of targeting ads explicitly based on special category data.[173] In 2024, several NGOs complained to the European Commission that LinkedIn's targeting function enabled advertisers to target ads (indirectly) based on special categories of data, and that this violated the DSA. After the European Commission formally requested information from the company, LinkedIn discontinued this targeting option.[174]

### 7.3  A ban on profiling-based advertising targeted at minors

Article 28(2) DSA concerns, in short, a ban on profiling-based advertising based on profiling that targets children.[175] Like the ban on profiling based on sensitive data, Article 28(2) DSA only applies to online platforms.[176]

The provision reads as follows.

> Providers of online platforms [177] shall not present advertisements on their interface based on profiling as defined in Article 4, point (4), of [the GDPR] using personal data of the

---

[173] Lex Zard, *Power and dignity: the ends of online behavioral advertising in the European Union*, PhD thesis University of Leiden, May 2024, https://hdl.handle.net/1887/3753619 accessed 26 May 2024, p. 170.

[174] European Commission, 'Statement by Commissioner Breton on steps announced by LinkedIn to comply with DSA provisions on targeted advertisement' (press release 7 June 2024) <https://digital-strategy.ec.europa.eu/en/news/statement-commissioner-breton-steps-announced-linkedin-comply-dsa-provisions-targeted-advertisement> accessed 29 October 2024. See also Patrick Corrigan (LinkedIn VP, Legal - Digital Safety), 'Welcoming the Digital Services Act' (update 7 June 2024), <https://www.linkedin.com/pulse/welcoming-digital-services-act-patrick-corrigan/> accessed 29 October 2024.

[175] See generally on this ban also: Zardiashvili (n 139) section 6.2.2.

[176] As noted previously (Section 4.6), for some types of adtech companies (such as ad networks) it is debatable whether they should be seen as online platforms.

[177] The DSA actually says: 'Providers of online platform', without the 's'. DSA, art. 28(2).

> recipient of the service when they are aware with reasonable
> certainty that the recipient of the service is a minor.

The ban applies when the platform knows 'with reasonable certainty that the recipient of the service is a minor'.[178] Article 28(3) adds that 'compliance with the obligations set out in this Article shall not oblige providers of online platforms to process additional personal data in order to assess whether the recipient of the service is a minor.' Platform providers are thus not obligated to collect extra data or use extra profiling to find out which users are minors. As mentioned in the DSA's preamble,[179] such extra data collection would be hard to square with the GDPR's data minimisation principle. At the same time, the DSA does not clarify under what circumstances the platforms should know that the recipient is a minor. Here, there is room for clarification in the announced guidelines under Article 28(4) DSA.[180]

The DSA's ban on targeting minors with advertising applies to advertising 'based on profiling' as defined in in the GDPR. This element of the ban's definition raises a similar question as the ban on profiling-based advertising using special categories of data. An online platform is probably still allowed to target ads to minors if the platform learns about the interests of a minor through other means than profiling. Suppose that Alice is 12 years old and discloses on her social network page that she likes horses. According to the letter of the DSA's ban, the platform is still allowed to target her with horse-related ads.

***Scope of the ban: a 'minor'***

---

[178] DSA, art 28(2).

[179] DSA, recital 71: 'In accordance with Regulation (EU) 2016/679 [the GDPR], notably the principle of data minimisation as provided for in Article 5(1), point (c), thereof, this prohibition [of Article 28(2) DSA] should not lead the provider of the online platform to maintain, acquire or process more personal data than it already has in order to assess if the recipient of the service is a minor. Thus, this obligation should not incentivize providers of online platforms to collect the age of the recipient of the service prior to their use.'

[180] See 'Commission launches call for evidence for guidelines on protection of minors online under the Digital Services Act' <https://digital-strategy.ec.europa.eu/en/news/commission-launches-call-evidence-guidelines-protection-minors-online-under-digital-services-act> accessed 20 December 2024.

The ban applies to profiling-based advertising to 'a minor'. The DSA nor its preamble define 'minor'. The words 'minor' and 'child' are not harmonised across the EU. Generally, people under 18 years are legally seen as minors in EU member states.[181] The United Nations Convention on the Rights of the Child also see people below 18 years old as children.[182] For these reasons and in the absence of contrasting arguments, we conclude that the term minor refers to someone who has not reached the age of 18 years.

***What does Article 28 DSA add to existing law?***

As we saw in the previous section, existing law (the GDPR and the ePrivacy Directive) already requires prior consent of the internet user for profiling-based advertising. For minors, the situation is more complicated. Roughly summarised, under the GDPR, minors cannot give valid consent; the parent should give consent instead.[183] EU member states have set different minimum consent ages, ranging from 13 to 16 years.[184] The company must make 'reasonable efforts' to verify that consent is given by the parent.[185] In sum, the GDPR, combined with the ePrivacy Directive, requires parental consent for profiling-based advertising targeted at children. Hence, the DSA provision on profiling-based advertising and children is stricter because it cannot be overridden by parental consent.

In addition, the Audiovisual Media Services Directive already prohibited, since 2018, profiling-based advertising based on personal data of minors.[186] That provision,

---

however, only applies to 'media service providers' as defined in that Directive. Such providers are, in short, people or organisations who have editorial responsibility for the choice of audiovisual content, such as videos.[187] The DSA's ban has a wider scope, because it applies to all online platforms.

## 8 Concluding thoughts

We analysed the main implications of the DSA for online advertising. We showed that the DSA's advertising rules also apply to non-commercial advertising, for instance by states, political parties and non-governmental organisations. Moreover, the DSA introduces several specific rules on online advertising.

The DSA bans online platforms from conducting profiling-based advertising that uses special categories of data (sometimes called sensitive data), such as data about somebody's religion, ethnicity and health. The DSA also bans profiling-based advertising that targets minors. Both types of advertising practices are only allowed under the GDPR after the data subject's consent or, for minors, the parent's consent. Compliance with, and enforcement of, the GDPR leave something to be desired, however. In this light, the DSA's restrictions on profiling-based advertising are a useful addition. The DSA also contains transparency requirements for online advertising. 'Very large' online platforms and search engines must develop advertising repositories, in which they show which ads were shown, with additional information.

Perhaps most importantly, we showed that the more general rules of the DSA also apply to advertisements and ad networks. Advertisements are a form of information, and thus subject to the general rules of the DSA (Section 3). For example, online platforms must

---

processed for commercial purposes, such as direct marketing, profiling and behaviourally targeted advertising.'

[187] Article 1(d), Audiovisual Media Services Directive, consolidated version https://eur-lex.europa.eu/legal-content/EN/TXT/HTML/?uri=CELEX%3A02010L0013-20181218.

offer transparency about the recommender systems for advertisements (Section 5.1) and offer a version of this recommender system that is not based on profiling (Section 7.1).

Moreover, we conclude that the DSA applies to some types of ad tech companies. More specifically, ad networks qualify as platforms, and some even as 'very large' platforms. The qualification of ad networks as online platforms leads to the application of the more general obligations in the DSA, such as obligations to work with trusted flaggers and address systemic risks (Sections 4.4 and 4.5). The relevance of these general obligations for advertising networks has been underexplored and deserves further research. We encourage the European Commission or regulators to clarify the concepts of 'online platform' and 'recipients' in the context of ad networks and other adtech companies. The scope of the two above-mentioned bans on profiling-based advertising could also use clarification. On thing is clear already. The DSA can have far-reaching effects for online advertising and adtech.

\* \* \*